\documentclass[useAMS,usenatbib]{mn2e}
\usepackage{graphicx}

\title[Accretion triggered oscillations in AK~Sco]{Evidence of accretion triggered oscillations in the pre-main-sequence interacting binary AK Sco}
\author[A. I. G\'omez de Castro and others]{Ana I G\'omez de Castro$^{1}$\thanks{E-mail:
aig@mat.ucm.es (AIG)}, Javier L\'opez-Santiago$^2$ and Antonio Talavera$^3$ \\
$^{1}$Fac. de CC Matem\'aticas, Universidad Complutense de Madrid, Plaza de Ciencias 3, 28040 Madrid, Spain\\
$^{2}$Fac. de CC. F\'{\i}sicas, Universidad Complutense de Madrid, Plaza de Ciencias 1, 28040  Madrid, Spain \\  
$^{3}$XMM-Newton Science Operations Centre, ESA, P.O. Box 78, 28691 Villanueva de la Ca\~nada, Spain }
\begin{document}

\date{Accepted . Received ; in original form }

\pagerange{\pageref{firstpage}--\pageref{lastpage}} \pubyear{}

\maketitle

\label{firstpage}

\begin{abstract}
Pre-main sequence (PMS) binaries are surrounded by circumbinary disks from which matter falls onto both components. The material dragged from the circumbinary disk flows onto each star through independent streams channelled by the variable gravitational field. The action of the bar-like potential is most prominent in high eccentricity systems made of two equal mass stars. AK Sco is a unique PMS system composed of two F5 stars in an orbit with e=0.47. Henceforth, it is an ideal laboratory to study matter infall in binaries and its role in orbit circularization.  In this letter, we report the detection of a 1.3mHz ultra low frequency oscillation in the ultraviolet light curve at periastron passage.  This oscillation last 7~ks being most likely fed by the gravitational energy released when the streams tails spiralling onto each star get in contact at periastron passage enhancing the accretion flow; this unveils a new mechanism for angular momentum loss during pre-main sequence evolution and a new type of interacting binary.

\end{abstract}

\begin{keywords}
binaries: close; binaries: spectroscopic; stars: pre-main-sequence; stars: individual: AK~Sco ; circumstellar matter, accretion: accretion disks
\end{keywords}

\section{Introduction}

The formation of binary stellar systems is not yet fully understood (Pringle 1991). A key problem is the abundance of close binaries that implies the action of efficient mechanisms for gravitational energy loss and angular momentum transport further than the tidal interaction between the stars. Star-disk interaction during the PMS evolution is thought to play a crucial role (Clarke \& Pringle, 1991).  Accretion disks can either take up or release angular momentum in the system that works like a multi-body gravitational system (Artymowicz \& Lubow, 1994; Bate \& Bonnell, 1997; Hanawa et al 2010).  The details depend on the mass ratio between the two stars and on the orbit eccentricity. Highly eccentric orbits favour the formation of spiral waves within the inner disk that do channel the flow as the accreting gas streams onto each star.  

AK Sco stands out among the PMS binaries by the high eccentricity of its orbit, e = 0.47; the two F5-type stars get as close as 11.3 stellar radii at periastron passage (see Table~1, for AK Sco main parameters). UV observations carried out with the Hubble Space Telescope (HST) unveiled the presence of a dense and extended structure around the stars with thermal properties alike those of the atmospheres/magnetospheres of late type PMS stars (G\'omez de Castro 2009). The structure extends up to 1.7 R$_*$ around each component and it is disturbed by a velocity field ($\sigma \simeq 100$ km s$^{-1}$) that exceeds significantly the sound speed in the UV radiating plasma, 30.5~km~s$^{-1}$. Thus, HST observations indicate that there are unresolved macroscopic motions in the dense and warm circumstellar material. Given the eccentricity of the orbit, we expected mass transfer to be significantly enhanced at periastron passage, forcing the response of the stellar magnetosphere to the accretion flow.

During PMS evolution, stellar magnetospheres play a key role as dissipative interfaces between the star and the disk. They do absorb and reprocess part of the angular momentum excess of the infalling material and channel it into open holes in the stellar magnetic configuration. Henceforth, magnetospheric radiation (ultraviolet and X-ray) is a very sensitive tracer of accretion. For this reason, we monitored AK~Sco with the XMM-Newton space telescope at three different phases, including the periastron passage (more details on the monitoring campaign can be found in G\'omez de Castro et al. 2012). However, it was only at periastron passage that an oscillation was detected at the rise of the light curve. The observations and data analysis are described in Section~2. In Section~3, the possible sources of the oscillation are analysed. A brief summary is provided in Section~4. 

\begin{table*}
 \centering
 \begin{minipage}{140mm}
  \caption{Main parameters of AK~Sco}
  \begin{tabular}{lll}
\hline
Property & Value & Reference\\
\hline
Projected semi-major axis & $a\sin i = 30.77 \pm 0.12$ R$_{\odot}$ & Andersen et al 1989 \\
Eccentricity & e= 0.47 & Andersen et al 1989, Alencar et al. 2003 \\
Orbital period & P=13.609 d & Andersen et al 1989, Alencar et al. 2003 \\
Inclination & $ i=65^o-70^o$ & Alencar et al. 2003 \\
Age & 10-30 Myrs & Alencar et al. 2003  \\
Spectral type & F5 & Alencar et al. 2003\\
Stellar Mass & $M_* = 1.35 \pm 0.07$M$_{\odot}$ & Alencar et al. 2003\\
Radius & $R_* = (1.59 \pm 0.35) R_{\odot}$ & Alencar et al. 2003 \\
Projected rotation velocity & $v \sin i = 18.5 \pm 1.0$ km s$^{-1}$ & Alencar et al. 2003 \\
Bolometric flux & $6.33\times 10^{-9}$ erg~s$^{-1}$cm$^{-2}$ & Andersen et al 1989\\
Extinction: A$_V$ & 0.5 mag & Manset et al. 2005 \\
Extinction: R & 4.3 & Manset et al. 2005 \\
Parallax & 9.73 mas & Van Leeuwen 2007 \\
\hline
\end{tabular}
\end{minipage}
\end{table*}

\section{Observations and data analysis}

AK~Sco observations were obtained on March 15th, 18th and 22nd,  2011.  They were performed at phases 0.0, 0.15 and 0.48, corresponding to observation identifications, ID 0651870201, ID 0651870301 and ID 0651870401, respectively.  Phase 0.0 corresponds to periastron passage. The exposure time of each observation was approximately 25 ks. In each observation, both the OM and the EPIC instruments were used to study the ultraviolet (UV) and X-ray evolution of the magnetospheric plasma, respectively. 

The OM was used in Image and Fast modes with the UVM2 filter 
(2310~\AA ). The AK Sco spectrum is dominated by magnetospheric radiation (mainly the Balmer continuum and the Fe II resonance UV multiplets)  in the wavelength range of this filter (2000 \AA\ - 2700 \AA ) thus the photospheric contribution is negligible (G\'omez de Castro \& Franqueira 1997). Each observation consisted of four exposures of 4400~s. Data were processed with the XMM-Newton Science Analysis System. We obtained light curves with a time resolution of one second, re-binned to 100 s to increase the signal to noise ratio. Fast oscillations were detected by OM  only from 4000~s to 12000~s during the observation performed at periastron passage; 
notice the three clear peaks between 6~ks and 9~ks in Figure~1. The peaks have the same structure
with a first heading maximum and a second weaker peak. The strength of the main peaks is above 2$\sigma$ in the light curve. 
The X-ray signal was too low to detect similar oscillations in that band.

\begin{figure}
 \includegraphics[width=84mm]{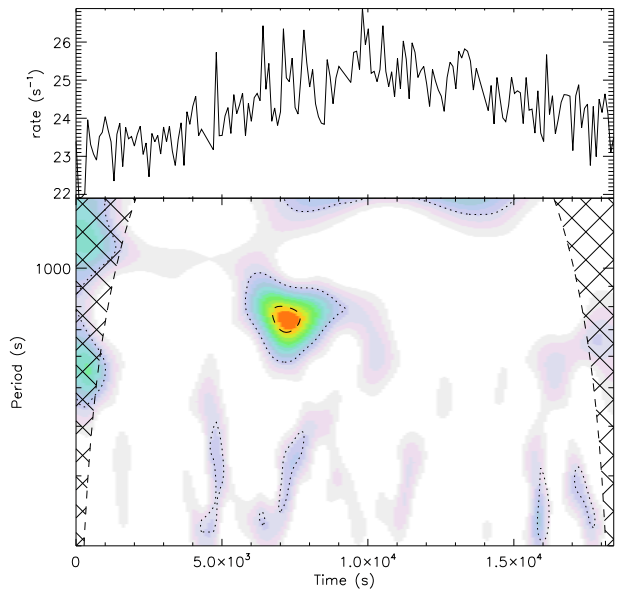}
 \caption{Upper panel:   OM light curve of AK Sco at periastron. Time binning is  100s . Lower panel: wavelet power spectrum of the upper light curve. The Y-axis is the period (P = 1/$\nu$) , with $\nu$ being the frequency of the power spectrum). Contour levels are represented for 1$\sigma$  and 2$\sigma$ significance levels. The dashed area represents periods for which results are not statistically reliable.}
\end{figure}

To analyse in more detail the oscillations, we used the method applied by Torrence \& Compo (1998) to the ''El Ni\~no" oscillation in the Atlantic Ocean. The analysis is based in the wavelet transform technique, a modified version of the Fourier analysis for the detection of quasi periodic signals. This method is essentially a form of windowed Fourier transform; it consists of convolving the original signal with a wavelet (or mother) function that is localised in both time and frequency domain. Examples of the use of this method in Astrophysics can be found in Bijaoui 2004, Welsh et al. 2006 or Mathioudakis et al 2006. 

We used the Morlet function as mother function (Mitra-Kraev et al. 2005), since it is especially well suited to detect non-stationary oscillations (Torrence \& Compo, 1998). The resultant power spectrum is also shown in Figure~1. The oscillation period was found to be 790$^{+200}_{-150}$~s, independently of  the temporal sampling of the OM light curve. The amplitude is 1.5$\pm$0.2 counts/s; about a 6\% oscillation over a 25 count/s signal level. In addition, we reconstructed the light curve by frequency domains. The frequency-filtered light curves are shown in Figure~2 and the sum of the three curves gives the observed OM light curve in Fig.~1. Retaining only the low frequency (large periods) permits to observe the large scale shape of the detected signal. Notice that the oscillation occurs in the middle of the rise. The middle panel shows the reconstructed light curve for frequencies around the peak of the wavelet power spectrum (see Fig.1). This light curve shows a clear modulation. For high frequencies, the reconstructed light curve shows basically white noise.

\begin{figure}
 \includegraphics[width=84mm]{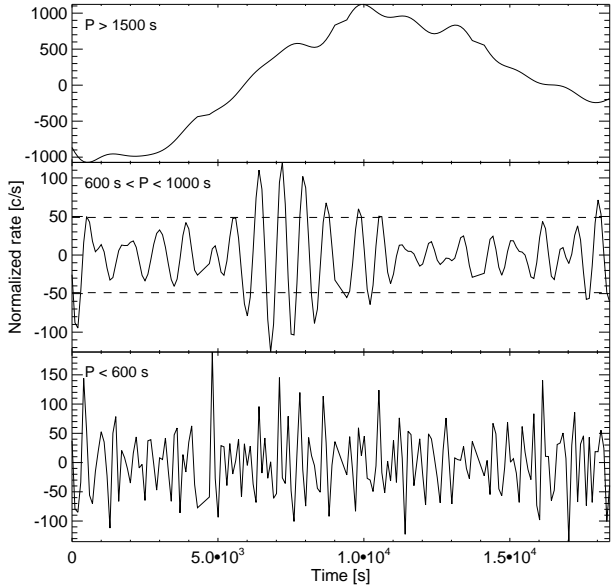}
 \caption{Reconstructed light curve after frequency filtering. The periods for the filtering are written in each panel. The scales are normalised for convenience. The low frequency (large periods) is shown in the top panel. The middle panel shows the reconstructed light curve for frequencies around the peak of the wavelet power spectrum (see Fig.1); the dashed lines mark the  +/-2$\sigma$ rms of the  light curve. The reconstructed light curve
for high frequencies is shown in the bottom panel. }
\end{figure}

\section{The source of the oscillation}

This ultra low frequency (ULF) oscillation represents a long wavelength mode excited around periastron passage and lasting only for few oscillations at the rise of the UV light curve. 

At periastron, the distance between both stars shrinks to 11.3~R$_*$ from a maximum of 31.4~R$_*$ distance at apastron passage. Studies of the structure and evolution of circumbinary disks predict the formation of circumstellar disks around each component (see e.g. Hanawa et al 2010).
Circumstellar disks are constrained within the Roche lobe thus matter falling from the circumbinary disk is being pulling towards the Lagrangian points L1, L2 and L3 feeding them. Matter orbits in the same direction in both circumstellar disks thus, when disks get in contact at periastron passage, a shear instability could be produced at the interface driving to angular momentum annihilation. An accretion outburst should be produced at this time. As shown in the light curve in Fig.s~1 and 2, the UV flux increases just before the periastron passage and the oscillation is observed in the rising of the light curve.

However, the precise mechanism producing the 790~s oscillation is unclear. 
One might expect that the ULF oscillation is tracing thermal oscillations in the inner border of the circumstellar disks. The avalanche of gas from the contact area between both disks, i.e. from the outer radius, would increase the density and opacity at the inner border of the circumstellar disks driving to thermal oscillations,  as often observed prior to the outburst in interacting  binaries (see e.g. Gaesincke et al. 2006). However, the ULF oscillation period is significantly larger than the thermal timescale, $\tau _{\rm th} \simeq  1/\alpha \Omega$ with  $\alpha$ the parameter prescribing transport in a thin circumstellar accretion disk and $\Omega $ the Keplerian frequency at the inner border of the disk.

This period, however, is close to that observed in the magnetosonic oscillations of flux tubes in the Sun (de Moortel 2009). Periods range from ~284~s in coronal loops (McEwan \& de Moortel, 2006) to 600-900~s in coronal plumes (Ofman et al. 1999). They are thought to be triggered by oscillations in the  loop foot-points that can either be impulsive (Nakariakov et al.  2004) or associated with solar p-modes  (McEwan \& de Moortel, 2006). 
Magnetoacustic oscillations with similar periods have only been detected in the very magnetically active star AT Mic apart from the Sun (Mitra-Kraev et al. 2005). But AK Sco cannot be considered as a very active star; it is an F5 star with very moderate X-ray flux. Henceforth, only a powerful external trigger could excite these oscillations to the level of making them detectable over its prominent UV excess. Only, the rapid enhancement of the accretion rate at periastron could play that role. 

During PMS evolution stellar magnetospheres extend to the inner border of the accretion disk since they rotate more slowly than the Keplerian disk. Infalling matter piles up at the magnetosphere boundary while its angular momentum excess is transferred to other scales or dissipated (Ghosh \& Lamb 1979, Goodson et al 1999, Romanova et al 2012). Magnetic flux tubes are expected to connect the stellar surface with this sheared interface layer (Favata et al. 2005). Rapid plasma penetration at the base of the loop, in the inner border of the accretion disk,  may excite oblique magnetosonic shock waves producing an impulsive flare.The additional energy filling the magnetic flux tube with hot plasma determines the relative amplitude of the oscillation thus $\Delta I/I = (c_S/v_A)^2 =0.06$, with $c_s$ the sound speed and $v_A$ the Alfv\'en speed in the loop. Assuming that the  thermal conditions within the loop are similar to those determined for the average corona from the X-ray  spectrum, the magnetic field in the loop is estimated to be B = 52.8 G, a modest value as expected for  an F5 type star. 
However, the natural oscillation period of the  magnetosonic modes is: $\tau  = 2L/jc_{T}$, with $L$ the length of the loop, $j$ the order of the harmonic and $(1/c_T)^2=  (1/c_S)^2+(1/v_A)^2$. Under coronal conditions, $\tau  = L/c_S$, for the first order mode. This corresponds to a natural loop length of 0.14R$_*$ for AK Sco, that seems to be very small for loops connecting the star to the inner border of its circumstellar disk. Non to mention, the high degree of synchronisation required to observe the oscillation simultaneously in both stars.

\begin{figure}
 \includegraphics[width=84mm]{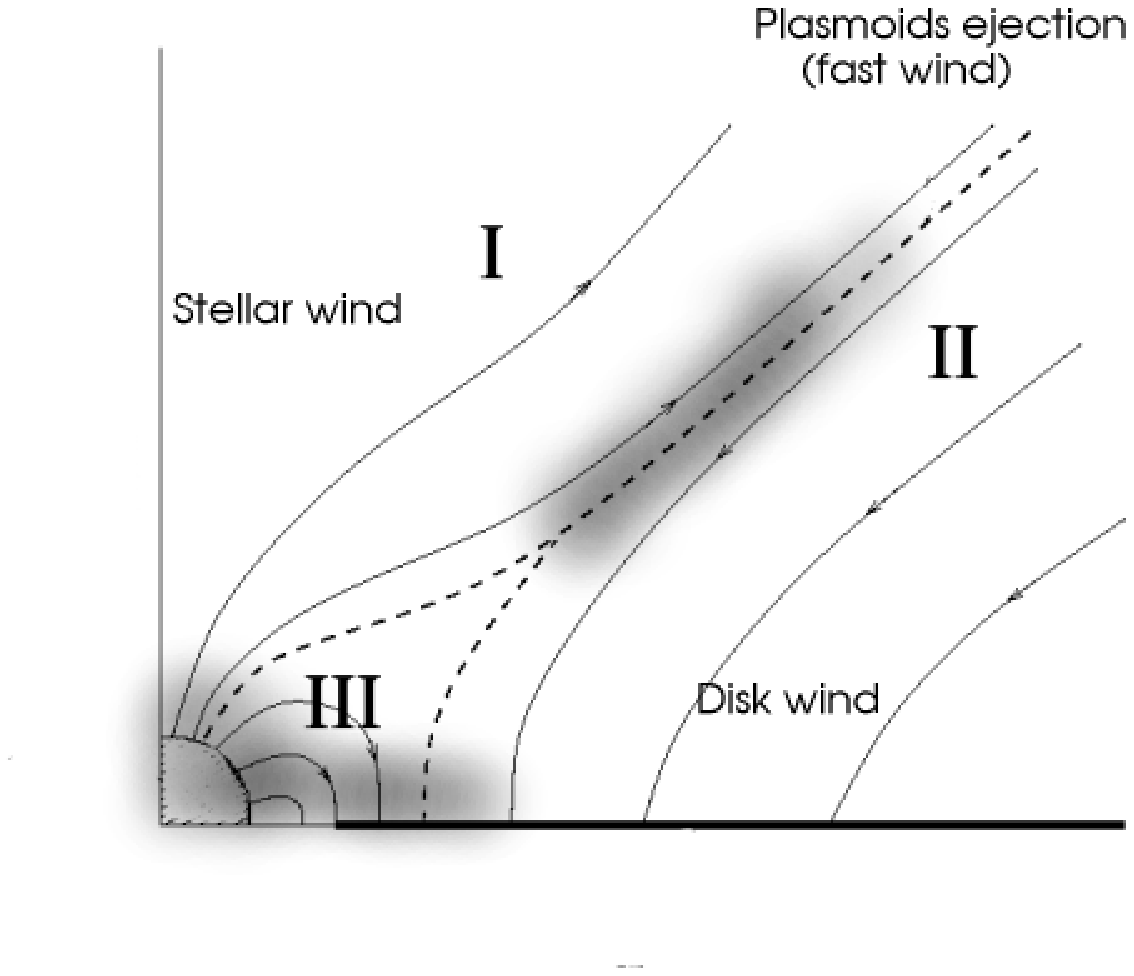}
 \caption{In PMS stars, the interaction between the stellar magnetic field and the
disk twists the stellar field lines due to the differential rotation.
The toroidal magnetic field generated out of the poloidal flux and
the associated pressure tend to push the field lines outwards,
inflating them, and eventually braking the magnetic link between
the star and the disk (boundary between regions I and II).
Three basic regions can be defined:
Region I dominated by the stellar wind, Region II dominated by
the disk wind and Region III dominated by stellar magnetospheric
phenomena. The dashed line traces the boundaries between this
three regions. The continuous lines indicate the topology
of the field and the shadowed areas represent regions where
magnetic reconnection events are likely to occur, producing
high energy radiation and particles (from G\'omez de Castro 2004). }
\end{figure}

The spiky nature of the oscillations suggest that they could be associated with bursts, typically triggered by reconnection events.
However, the large amplitude of the outburst (6\%) indicates that there is an external trigger as accretion. 
In PMS stars, the stellar field is assumed to be anchored in the inner part of the disk creating a sheared layer between the 
rigid body rotation of the star and the Keplerian rotation of the disk. This interaction has a profound influence on the star
and the accretion flow but also acts as a dynamo that transform part of the angular momentum excess in the inner disk
into magnetic field amplification that self-regulates through quiescent periods of field building-up and eruptions when the energy excess is released by magnetic reconnection in the current layer (see Figure 3). There are however, a great number of uncertainties in the way the system self-regulates that, in turn, depends on magnetic diffusivity (see for instance, Goodson \& Winglee, 1999).
Typical diffussivities range from $\eta =  92.34$~m$^2$s$^{-1}$ in the dense magnetospheric components radiating the ultraviolet to 
$\eta = 0.09$~m$^2$s$^{-1}$ in the low density, X-ray radiating plasma. 
If the ULF oscillation was associated with quasi-periodic oscillations in reconnection events within the disk-star boundary layer, the inferred physical scales, $L_0$, would be small since $L_0 = (\eta 790 s )^{1/2}$ and compatible with the expected thickness of the layer.

\section{Conclusions}

We report the detection of an ULF oscillation in the UV light curve of the PMS binary AK Sco at the rise of the light curve. This is the first detection of an accretion triggered oscillation in a PMS binary and makes AK~Sco an ideal laboratory to study angular momentum evolution in these systems as well as close binary formation. This work unveils a new type of interactive binary where outburst are driven by angular momentum loss between circumstellar disks. However, further observations are required to identify the physical nature of the oscillation. 
The frequency, timing and strength of the oscillation at various wavelengths and thermal regimes, needs to be investigated, as well as the 
repeatability of the phenomenon. The detection of the ULF oscillation opens up the way for detailed studies on the properties of PMS magnetospheres and their interaction with the accretion flow.

\section*{Acknowledgments}

We thank our anonymous referee for insightful comments on the possible source of the ULF oscillation.
We thank our colleagues D. Bisikalo and A. Sytov for enlightening discussions on the dynamics of the AK~Sco system. AIGdC acknowledges financial support by the Spanish Government under grant AYA2011-29754-C03-01.  JL-S acknowledges financial support by the Spanish Government under grants AYA2011-30147-C03-02 and AYA2011-29754-C03-03. JL-S also  thanks project AstroMadrid (S2009/ESP-1496) for partial support.

\label{lastpage}


\begin{thebibliography}{99}

\bibitem[]{}
Alencar, S.H.P. et al.,  2003, A\&A, 409, 1037-1053 
\bibitem[]{}
Andersen, J., Lindgren, H., Hazen, M.L., Mayor, M., 1989, A\&A, 219, 142-150 
\bibitem[]{}
Artymowicz, P.,  Lubow, S.H. 1994, ApJ, 421, 651-667 
\bibitem[]{}
Bate, M.R., Bonnell, I.A, 1997,  MNRAS, 285, 33-48
\bibitem[]{}
Bijaoui, A.  2004,  Wavelets in Physics, Cambridge University Press, ISBN -13: 9780521533539, 77-114 
\bibitem[]{}
Clarke, C. J., Pringle, J.E., 1991, MNRAS, 249, 588-595 
\bibitem[]{}
De Moortel, I., 2009,  Space Science Rev., 149, 65-81 
\bibitem[]{}
Favata, F., Flaccomio, E., Reale, F., et al. 2005, ApJSS,   160,  469-502 
\bibitem[]{}
Gaensicke, B., de Martino, D.,  Marsh, T.R. et al. 2006, ApSS, 303, 177
\bibitem[]{}
Ghosh, P., Lamb, F.K,  1979,  ApJ, 232, 259
\bibitem[]{}
G\'omez de Castro, A.I., 2004, ApSS, 292, 561
\bibitem[]{}
G\'omez de Castro, A.I. 2009, ApJ, 698, L109-L111 
\bibitem[]{}
G\'omez de Castro, A.I., Franqueira, M., 1997, IUE-ULDA Guide n. 8 to T Tauri Stars, European Space Agency  Publications SP-1205, 1-349
\bibitem[]{}
G\'omez de Castro, A.I., L\'opez-Santiago, J., Talavera, A. et al., 2012, ApJ, submitted.
\bibitem[]{}
Goodson, A.P., Boehm, K-H, Winglee, R.M., 1999, ApJ, 524, 142
\bibitem[]{}
Goodson, A.P., Winglee, R.M., 1999, ApJ, 524, 159
\bibitem[]{}
Hanawa, T.,  Ochi, Y.,  Ando, K.,  2010,  ApJ, 708, 485-497 
\bibitem[]{}
Manset, N., Bastien, P., Bertout, C., 2005, AJ, 129, 480
\bibitem[]{}
Mathioudakis, M., Bloomfield, D.S., Jess, D.~B. et al., 2006, A\&A, 456, 323
\bibitem[]{}
McEwan, M. P., de Moortel, I.,  2006, A\&A,  448, 763-770 
\bibitem[]{}
Mitra-Kraev, U., Harra, L.K., Williams, D. R., Kraev,E., 2005,  A\&A, 436, 1041-1047 
\bibitem[]{}
Nakariakov, V.M., Tsiklauri, D., Kelly, A., et al., 2004, A\&A, 414, L25-L28 (2004) 
\bibitem[]{}
Ofman, L., Nakariakov, V.M.,  DeForest, C.E., 1999, ApJ,  514, 441-447 
\bibitem[]{}
Pringle, J.E., 1991, {\it The Physics of Star Formation and Early Stellar Evolution}, NATO Advanced Science Institutes (ASI) Series C, Dordrecht: Kluwer, 1991, edited by Charles J. Lada and Nikolaos D. Kylafis., 342, 437 
\bibitem[]{}
Romanova, M.M., Ustyugova, G.V., Koldoba, A.V., Lovelace, R.V.E., 2012, MNRAS, 421, 63
\bibitem[]{}
Torrence, C.,  Compo, G.P. 1998,  Bulletin of the American Meteorological Society, 79, 61- 78
\bibitem[]{}
van Leeuwen, F., 2007, A\&A, 474, 653-664 
\bibitem[]{}
Welsh, B.Y., Wheatley, J., Browne, S.E. et al. 2006,  A\&A, 458, 921


\end{thebibliography}
\end{document}